\documentclass[superscriptaddress,showpacs,amsmath,amssymb,twocolumn]{revtex4-1}
\usepackage{amsmath, amsthm, amssymb, bm, braket}
\usepackage[dvips]{graphicx}
\usepackage{color}
\usepackage{wrapfig}
\usepackage{fancybox}

\begin{document}

\title{Finite amplitude method applied to giant dipole resonance in heavy rare-earth nuclei}

\author{Tomohiro Oishi}
\email[Electronic address: ]{tomohiro.t.oishi@jyu.fi} 
\affiliation{Helsinki Institute of Physics, P.O. Box 64, 
FI-00014 University of Helsinki, Finland}
\affiliation{Department of Physics, P.O. Box 35 (YFL), 
University of Jyvaskyla, FI-40014 Jyvaskyla, Finland} 

\author{Markus Kortelainen}
\affiliation{Department of Physics, P.O. Box 35 (YFL), 
University of Jyvaskyla, FI-40014 Jyvaskyla, Finland}
\affiliation{Helsinki Institute of Physics, P.O. Box 64, 
FI-00014 University of Helsinki, Finland}

\author{Nobuo Hinohara}
\affiliation{Center for Computational Sciences, University of Tsukuba, Tsukuba 305-8577, Japan}
\affiliation{FRIB Laboratory, Michigan State University, East Lansing, Michigan 48824, USA}


\renewcommand{\figurename}{FIG.}
\renewcommand{\tablename}{TABLE}

\newcommand{\bi}[1]{\ensuremath{\boldsymbol{#1}}}
\newcommand{\unit}[1]{\ensuremath{\mathrm{#1}}}
\newcommand{\oprt}[1]{\ensuremath{\hat{\mathcal{#1}}}}
\newcommand{\abs}[1]{\ensuremath{\left| #1 \right|}}

\newcommand{\crc}[1] {c^{\dagger}_{#1}}
\newcommand{\anc}[1] {c_{#1}}
\newcommand{\crb}[1] {a^{\dagger}_{#1}}
\newcommand{\anb}[1] {a_{#1}}

\def \beq{\begin{equation}}
\def \eeq{\end{equation}}
\def \beqa{\begin{eqnarray}}
\def \eeqa{\end{eqnarray}}

\def \bir{\bi{r}}
\def \ubir{\bar{\bi{r}}}
\def \bip{\bi{p}}
\def \ubip{\bar{\bi{r}}}

\begin{abstract}
{\noindent {\bf Background:} 
The quasiparticle random phase approximation (QRPA), 
within the framework of the nuclear density functional theory (DFT), 
has been a standard tool to access the collective excitations of the atomic nuclei. 
Recently, finite amplitude method (FAM) has been developed, 
in order to perform the QRPA calculations efficiently without any 
truncation on the two-quasiparticle model space. \\
{\bf Purpose:} 
We discuss the nuclear giant dipole resonance (GDR) in heavy rare-earth isotopes, 
for which the conventional matrix diagonalization of the QRPA is numerically demanding. 
A role of the Thomas-Reiche-Kuhn (TRK) sum rule enhancement factor, 
connected to the isovector effective mass, is also investigated. \\
{\bf Methods:} 
The electric dipole photoabsorption cross section was calculated 
within a parallelized FAM-QRPA scheme. We employed the Skyrme energy density
functional self-consistently in the DFT calculation for the ground states and
FAM-QRPA calculation for the excitations.\\
{\bf Results:} 
The mean GDR frequency and width are mostly reproduced with the FAM-QRPA, 
when compared to experimental data, although some deficiency is observed with isotopes heavier than erbium.
A role of the TRK enhancement factor in actual GDR strength is clearly shown: 
its increment leads to a shift of the GDR strength to higher-energy region, without 
a significant change in the transition amplitudes. \\
{\bf Conclusions:} 
The newly developed FAM-QRPA scheme shows a remarkable efficiency, which 
enables to perform systematic analysis of GDR for heavy rare-earth nuclei. 
Theoretical deficiency of photoabsorption cross section
could not be improved by only adjusting the TRK enhancement factor, suggesting
the necessity of beyond the self-consistent QRPA approach, and/or a more systematic optimization of the EDF parameters.}
\end{abstract}

\maketitle

\section{Introduction} \label{Sec:intro}
Collective excitations of the atomic nuclei reflects various properties 
of the nuclear structure and the underlying interaction between nucleons. 
Their macroscopic or microscopic description has been 
a major subject in the nuclear theory~\cite{1944Mig,1948GT,1950SJ,Harakeh-Woude,80Ring,94BB}. 
Recently, self-consistent mean-field models, based on the nuclear 
density functional theory (DFT), have been intensively applied to the 
collective excitations in heavy open-shell nuclei, where {\it ab initio} 
models are still not computationally feasible. 

The giant dipole resonance (GDR) is a noticeable phenomenon generated 
by the electric dipole excitation. 
It is basically understood as a collective oscillation of all the 
neutrons against all the protons, occupying a major part 
of the nuclear giant resonances \cite{1944Mig,1948GT,1950SJ,Harakeh-Woude}. 
The GDR plays an essential role in the nuclear photoabsorption reaction, 
determining the centroid energy and width of the cross section. 
Nuclear photo-absorption reaction impacts also on the dynamics
of various astrophysical scenarios~\cite{2007Arnould}.
Therefore, GDR can provide a good testing ground for DFT-based theories to describe 
the nuclear collectivity, as well as the relevant physical 
properties of finite and infinite nuclear systems. 
For example, the centroid energy of GDR, which is well approximated 
as $\hbar \omega \cong 80A^{-1/3}$ MeV for spherical nuclei, 
can be connected to the symmetry energy in the infinite nuclear matter, 
which is an important pseudo-observable used to determine the 
parameters of the nuclear energy density functional (EDF)~\cite{94BB,1995Rein,1999Rein}. 
The wide spread of the GDR in neutron-rich nuclei \cite{RevModPhys.47.713}
has been understood to originate from the ground state deformation, which has been well reproduced with the modern nuclear EDFs \cite{arteaga:034311,PhysRevC.86.034328,2011Yoshida,2013Yoshida,2008Rein}. 
Also, the pairing part of the nuclear EDF has been expected to play a 
significant role in the low-lying dipole excitations of exotic nuclei \cite{2005Matsuo,2011Oishi,2013Dinh,PhysRevC.90.024303,PhysRevC.92.049902}. 

A commonly used DFT-based approach to address collective nuclear excitations is done in the
framework of linear response theory, with random-phase approximation (RPA). 
By taking the pairing correlations into account, 
the RPA is extended to the quasiparticle random-phase approximation (QRPA), 
which has been conventionally treated in the matrix formulation \cite{80Ring,2003Dean}. 
A fully self-consistent calculation within the matrix QRPA could be, 
however, numerically demanding due to the large size of QRPA matrices. 
Especially, in the case where the spherical symmetry is broken, 
one often needs to employ
an additional truncation on the two-quasiparticle model space 
in order to reduce the numerical cost \cite{2010TE,2011TE,2011Yoshida,arteaga:034311}.
Another approach to reduce the computational cost of the QRPA is the separable approximation for the residual interaction \cite{PhysRevC.66.044307,PhysRevC.74.064306,doi:10.1142/S0218301308009586,2008Rein}.
Such a truncation or approximation, however, may invoke the spurious excitations due to the 
broken self-consistency between the Hartree-Fock-Bogoliubov (HFB) ground state
and the QRPA solution.

The finite-amplitude method (FAM) provides an alternative way to 
solve the QRPA problem with a significantly reduced computational cost. 
With this method, the QRPA linear response problem is solved iteratively, by
circumventing actual calculation and diagonalization of the QRPA matrix. 
FAM was originally developed for a computation of the RPA strength function, and soon after 
it was expanded to cover the QRPA problem within spherical symmetry \cite{2007Naka,2011AN}. 
In Ref.~\cite{2011Markus}, FAM-QRPA was implemented into the axially symmetric
Skyrme-HFB solver, based on the harmonic oscillator basis.
Up to the date, FAM has been implemented also to 
the axially symmetric coordinate-space HFB solver~\cite{2014Pei} and 
to the relativistic mean-field framework~\cite{2013Liang,2013Nik}. 
Various applications of the FAM include 
descriptions of giant and pygmy dipole excitations \cite{2009Ina,2011Ina}, 
efficient computation of the QRPA matrix elements \cite{2013AN}, and 
evaluation of beta-decay rates, including the proton-neutron pairing correlations~\cite{2014Mika,2015Mika}. 
The contour integration technique of FAM-QRPA was developed to
describe individual QRPA modes \cite{2013Hino} and 
for a fast calculation of the energy-weighted sum rules \cite{2015Hino}. 
In addition to FAM, the iterative Arnoldi method presents an alternative method
to solve the QRPA problem iteratively~\cite{2010Gill}. It was also applied to 
the multipole excitations with pairing correlations \cite{2012Gill_1,2012Gill_2}. 

This article is devoted to FAM-QRPA methodology applied to the GDR of the heavy rare-earth nuclei,
within the Skyrme EDF framework. Due to the open-shell nature of these nuclei, pairing and
deformation properties must be taken into account in systematic study.
We do not assume any truncation of the two-quasiparticle model space in the QRPA, 
nor the Bardeen-Cooper-Schrieffer (BCS) approximation for the pairing, but keep a full self-consistency between the HFB and QRPA.
To check the validity of the FAM-QRPA, the results are compared with several sets of experimental data.
We also investigate the impact of the Thomas-Reiche-Kuhn (TRK) sum rule enhancement factor on 
the isovector dipole excitation. Because the TRK sum rule is independent on theoretical models and only the 
enhancement factor (or equivalently, the isovector effective mass) includes the information on the nuclear structure, 
the energy-weighted sum rule of GDR is an important quantity which 
reflects the properties of EDFs \cite{80Ring,1995Rein,1999Rein}.
The sensitivity of GDR to the isovector effective mass is also discussed. 

We introduce the basic formalism of the Skyrme EDF and FAM-QRPA in the next section. 
The results are presented and discussed in Sec.~\ref{sec:res}. 
Finally, we summarize this article in Sec.~\ref{sec:sum}. 

\section{Formalism} \label{Sec:formalism}
As a starting point, our HFB calculations were done in the Skyrme EDF framework.
In order to write the Skyrme energy density, it is convenient to introduce the isoscalar and isovector local densities
\beq
 \rho_0(\bir)=\rho_{\rm n}(\bir)+\rho_{\rm p}(\bir),~~~\rho_1(\bir)=\rho_{\rm n}(\bir)-\rho_{\rm p}(\bir), 
\eeq
where $\rho_{\rm n}$ and $\rho_{\rm p}$ are the neutron and proton densities. 
With these densities, the Skyrme energy density for the particle-hole (ph) channel 
reads as
\beqa
 \mathcal{E}^{\rm Skyrme}
 &=& \sum_{t=0,1} \left[ \mathcal{E}^{\rm even}_t + \mathcal{E}^{\rm odd~}_t \right], \\
 \mathcal{E}^{\rm even}_t
 &=& C^{\rho \rho}_t[\rho_0]\rho^2_t + C^{\rho \Delta \rho}_t\rho_t \Delta \rho_t + C^{\rho \tau}_t\rho_t \tau_t \nonumber \\
 & & + C^{\rho \nabla J}_t\rho_t \nabla \cdot \bi{J}_t + C^{JJ}_t\bi{J}^2_t, \\
 \mathcal{E}^{\rm odd~}_t
 &=& C^{ss}_t[\rho_0]\bi{s}^2_t + C^{s \Delta s}_t\bi{s}_t \cdot \Delta \bi{s}_t + C^{sT}_t\bi{s}_t \cdot \bi{T}_t \nonumber \\
 & & + C^{s \nabla j}_t\bi{s}_t \cdot (\nabla \times \bi{j}_t) + C^{jj}_t\bi{j}^2_t,
\eeqa
where $t=0~(1)$ indicates the isoscalar (isovector) components. 
The time-even part $\mathcal{E}^{\rm even}$, is a functional of the 
local density $\rho$, kinetic density $\tau$, and spin-orbit density $\bi{J}$, 
whereas 
the time-odd part $\mathcal{E}^{\rm odd}$ is expressed with 
the spin density $\bi{s}$, current density $\bi{j}$, and kinetic-spin density $\bi{T}$. 
The detailed formulation of these quantities can be found in, {\it e.g.}, Refs.~\cite{1996Jacek,2009Klupfel}. 
The coupling coefficients, $C^{\rho \rho}_0$, etc., are uniquely 
related to the well-known $(t,x)$ parameterization of the Skyrme force \cite{2009Klupfel,2004Jacek}. 
Also, some of coupling constants can be connected to the properties of the 
symmetric or asymmetric nuclear matter, which are useful pseudo-observables for the 
optimization purposes of the Skyrme EDF parameters. 
These pseudo-observables can be treated as alternative EDF input parameters instead of coupling constants~\cite{2010Markus}. 

In HFB calculation for the ground state of even-even nuclei, a time-reversal symmetry is usually assumed, 
and hence, the time-odd part of the functional does not make contribution to the HFB solution. 
On the other side, when the time-reversal symmetry becomes broken, 
like in the case of QRPA, the time-odd part becomes active. If we start from the original Skyrme force, 
the consequent time-odd part of the EDF has a unique correspondence to the time-even part. 
In other words, when we fix the coupling coefficients in the time-even part, 
those in the time-odd part should be automatically determined.
In the EDF framework, however, a further generalization can be considered:
one may treat the time-odd coefficients independently from the time-even ones.
In this work, time-odd part is determined as in the case of Skyrme force.
For Coulomb energy density, the direct term is treated in a usual manner and for the exchange part we employ 
Slater approximation.

For the particle-particle (pp) channel, which describes nuclear pairing correlations, 
we employ a functional of the density-dependent delta pairing (DDDP) energy density. 
That is,
\beq
 \mathcal{E}^{\rm pair} = \sum_{q={\rm n,p}} \frac{V^{\rm pair}_q}{2} \left[ 1 - \zeta \frac{\rho_0(\bir)}{\rho_c} \right] \tilde{\rho}^2_q(\bir), 
\eeq
where $\tilde{\rho}$ is the local pairing density and $\rho_c = 0.16$ fm$^{-3}$ 
is the nuclear saturation density. 
In this article, a mixed DDDP ($\zeta=1/2$) is adopted. 
The pairing strengths $V_q^{\rm pair}$ will be adjusted in Sec. \ref{sec:res}. 

\subsection{Finite amplitude method}
The detailed formulation of FAM-(Q)RPA can be found in Refs.~\cite{2007Naka,2011AN,2011Markus,2013Hino}. 
We briefly follow these works to arrange the formalism necessary in this work. 
First, we assume an external time-dependent field, inducing a polarization on the HFB ground state. 
This external field is
\beqa
 && \oprt{F}(t) = \eta \left[ \oprt{F} e^{-i\omega t} + \oprt{F}^{\dagger} e^{i\omega t} \right], \nonumber \\
 && \oprt{F} = \frac{1}{2} \sum_{\mu \nu} 
    \left[ F^{20}_{\mu \nu}(\anb{\nu} \anb{\mu})^{\dagger} + F^{02}_{\mu \nu}\anb{\nu} \anb{\mu} \right],
\eeqa
where $\crb{\mu}$ and $\anb{\nu}$ are the quasiparticle creation and annihilation 
operators, respectively, and $\eta$ is an infinitesimal real parameter. 
In this article, $\oprt{F}$ is assumed to be independent of $\omega$, and 
restricted to have the form of the one-body operator. 
That is, 
\beq
 \oprt{F} = \sum_{kl}f_{kl} \crc{k} \anc{l},
\eeq
where $\crc{k}$ and $\anc{l}$ are the original particle creation and annihilation operators. 
The expressions of $F^{20}_{\mu \nu}$ and $F^{02}_{\mu \nu}$ in terms of 
the Bogoliubov transformation can be found {\it e.g.} in Refs.~\cite{80Ring,2011AN}. 

Time evolution of quasiparticles is described by the time-dependent HFB equation, 
\beq
 i\frac{\partial}{\partial t} \anb{\mu}(t) = \left[ \oprt{H}(t)+\oprt{F}(t), \anb{\mu}(t) \right], \label{eq:TDHFB}
\eeq
where the deviation from the static HFB solution is represented as 
\beqa
 && \anb{\mu}(t) = e^{iE_{\mu}t} \left[ \anb{\mu} + \delta \anb{\mu}(t) \right], \nonumber \\
 && \delta \anb{\mu}(t) = \eta \sum_{\nu} \crb{\nu} \left[ X_{\nu \mu}(\omega) e^{-i\omega t} + Y^*_{\nu \mu}(\omega) e^{i\omega t} \right].
\eeqa
The quantities needed to obtain the multipole transition strength are the FAM amplitudes, 
$X_{\nu \mu}(\omega)$ and $Y_{\nu \mu}(\omega)$, at the excitation energy $\omega$. 
Since the external field induces density oscillations atop of the static HFB density, 
the self-consistent Hamiltonian also contains an induced oscillation: 
$\oprt{H}(t)=\oprt{H}_{\rm HFB}+\delta \oprt{H}(t)$, where 
\beqa
 && \delta \oprt{H}(t) = \eta \left[ \delta \oprt{H} e^{-i\omega t} + \delta \oprt{H}^{\dagger} e^{i\omega t} \right], \nonumber \\
 && \delta \oprt{H} = \frac{1}{2} \sum_{\mu \nu} 
    \left[ \delta H^{20}_{\mu \nu}(\omega) (\anb{\nu} \anb{\mu})^{\dagger} + \delta H^{02}_{\mu \nu}(\omega) \anb{\nu} \anb{\mu} \right]. 
\eeqa
By solving Eq.~(\ref{eq:TDHFB}) up to the first order in $\eta$ yields so-called FAM equations 
\beqa
 \left[ E_{\mu} + E_{\nu} - \omega \right] X_{\mu \nu} (\omega) + \delta H^{20}_{\mu \nu} (\omega) &=& -F^{20}_{\mu \nu}, \nonumber \\
 \left[ E_{\mu} + E_{\nu} + \omega \right] Y_{\mu \nu} (\omega) + \delta H^{02}_{\mu \nu} (\omega) &=& -F^{02}_{\mu \nu}. \label{eq:FAM_eq}
\eeqa
It is worthwhile to note that, by using the expressions of 
$\delta H^{20}_{\mu \nu}(\omega)$ and $\delta H^{02}_{\mu \nu}(\omega)$ in terms of 
$X_{\mu \nu}(\omega)$ and $Y_{\mu \nu}(\omega)$, 
one can transform Eq.~(\ref{eq:FAM_eq}) into the matrix form of
\beq
 \left[ \left( \begin{array}{cc} A & B \\ B^* & A^* \end{array} \right)
 - \omega \left( \begin{array}{cc} {\bf 1} & 0 \\ 0 & -{\bf 1} \end{array} \right) \right]
 \left( \begin{array}{c} X(\omega) \\ Y(\omega) \end{array} \right)
 = - \left( \begin{array}{c} F^{20} \\ F^{02} \end{array} \right), \label{eq:MFAM}
\eeq
where $A$ and $B$ are the well-known QRPA matrices \cite{80Ring}. 
Notice that Eq.~(\ref{eq:MFAM}) yields the standard matrix form of QRPA when external field is set to zero. 
The solution of Eq.~(\ref{eq:MFAM}), however, would require to compute the QRPA matrices $A$ and $B$
which generally have large dimension, leading to a substantial CPU time requirement.
The essential trick of the FAM-QRPA is to keep Eq.~(\ref{eq:FAM_eq}), and to solve the FAM amplitudes iteratively 
with respect to the response of the self-consistent Hamiltonian. 
This allows to circumvent the large numerical cost of matrix QRPA.

The response of the self-consistent Hamiltonian, 
$\delta H^{20}_{\mu \nu}(\omega)$ and $\delta H^{02}_{\mu \nu}(\omega)$, 
can be expressed in terms of the induced fields: 
\beqa
 \delta H^{20}_{\mu \nu}(\omega) &=& \left\{ U^{\dagger} \delta h(\omega) V^* - V^{\dagger} \delta h(\omega)^{T} U^* \right. \nonumber \\
 && \left. -V^{\dagger} \overline{\delta\Delta}(\omega)^* V^* + U^{\dagger} \delta \Delta(\omega) U^* \right\}_{\mu\nu}, \nonumber \\
 \delta H^{02}_{\mu \nu}(\omega) &=& \left\{ U^T \delta h (\omega)^T V - V^T \delta h(\omega) U \right. \nonumber \\
 && \left.  -V^T \delta \Delta(\omega) V + U^T \overline{\delta \Delta}(\omega)^* U \right\}_{\mu\nu}
\eeqa 
with the well-known HFB matrices, $U$ and $V$. 
Originally the induced FAM-QRPA fields, 
$\delta h$, $\delta \Delta$ and $\overline{\delta \Delta}$, were 
calculated by applying numerical functional derivatives. 
In Ref.~\cite{PhysRevC.92.051302}, on the other side, 
these fields were obtained through explicit linearization of the Hamiltonian, in order not to 
mix the densities with different magnetic quantum numbers $K$. 
Thanks to this explicit linearization, the infinitesimal parameter $\eta$
is no longer needed, and the induced fields can be formulated in 
the similar manner as the HFB fields. 
That is, 
$\delta h(\omega) = h' [\rho_f,\kappa_f,\overline{\kappa}_f]$, 
$\delta \Delta(\omega) = \Delta' [\rho_f,\kappa_f]$ and 
$\overline{\delta \Delta}(\omega) = \Delta' [\overline{\rho}_f,\overline{\kappa}_f]$, 
where $h'$ and $\Delta'$ are the linearized fields with respect of perturbed densities. 
These densities can be expressed as 
\beqa
 \rho_f(\omega) &=& +UX(\omega)V^T+V^*Y(\omega)^TU^{\dagger}, \nonumber \\
 \overline{\rho}_f(\omega) &=& +V^*X(\omega)^{\dagger}U^{\dagger}+UY(\omega)^*V^T, \nonumber \\
 \kappa_f(\omega) &=& -UX(\omega)^TU^T-V^*Y(\omega)V^{\dagger}, \nonumber \\
 \overline{\kappa}_f(\omega) &=& -V^*X(\omega)^*V^{\dagger}-UY(\omega)^{\dagger}U^T. 
\eeqa
The procedures that provide $h$ and $\Delta$ for the HFB solution 
can be also utilized for the linearized fields, $h'$ and $\Delta'$, 
with a minor modification. 
For an iterative solution of the FAM amplitudes, the Broyden method was utilized 
to obtain a rapid convergence \cite{1988J_Broy,2008B_Broy}. 

By using the FAM-QRPA amplitudes obtained through the iteration, the multipole transition strength 
distribution is expressed as 
\beqa
 \frac{dB(\oprt{F};\omega)}{d\omega} &\equiv & \sum_{i>0} \abs{\Braket{i|\oprt{F}|0}}^2 \delta(\omega-\Omega_i) \nonumber  \\
 &=& -\frac{1}{\pi} {\rm Im} S(\oprt{F};\omega), 
 \eeqa
where $i>0$ denotes the summation over the states with positive QRPA energies $\Omega_i>0$, and the response function is given by
$S(\oprt{F};\omega) = {\rm tr} [f \rho_f]$ \cite{2011AN,PhysRevC.92.051302}. 
In order to prevent the FAM-QRPA strength from diverging at $\omega=\Omega_i$, 
we employ a small imaginary part in the energy,
$\omega \rightarrow \omega_{\gamma}=\omega+i\gamma$, 
corresponding to a Lorentzian smearing of $\Gamma=2\gamma$ \cite{2011AN}. 
The explicit formulation of this smeared strength can be found 
in Ref.~\cite{2013Hino}: 
\beq
 S(\oprt{F};\omega_{\gamma}) = -\sum_{i>0} 
 \left( \frac{\abs{\Braket{i|\oprt{F}|0}}^2}{\Omega_i-\omega-i\gamma} + \frac{\abs{\Braket{0|\oprt{F}|i}}^2}{\Omega_i+\omega+i\gamma} \right).
\eeq
The contour integration technique is worth to be mentioned: discrete QRPA amplitudes 
or various multipole sum rules can be obtained from $S(\oprt{F};\omega_{\gamma})$ with 
a suitable selection of the integration contour on a complex ($\omega,\gamma$)-plane~\cite{2013Hino,2015Hino}. 

We use following external fields to compute the electric isovector dipole (IVD) 
strength $dB(\oprt{D}_K;\omega)/d\omega$,
\beqa
 \oprt{F} &=& \oprt{D}_{K}~~~(K=0,\pm 1), \nonumber \\
 \oprt{D}_{K} &=& e \frac{NZ}{A} \left[ \sum_{i \in N} \frac{-1}{N} D_K(\bir_i) + \sum_{j \in Z} \frac{1}{Z} D_K(\bir_j) \right], \label{eq:F_IVD} 
\eeqa
with $D_K(\bir) = rY_{1K}(\ubir)$. 
In actual calculation, we replace this operator as 
\beq
 D_K \rightarrow D^+_K = (D_K+D_{-K})/\sqrt{2-\delta_{0K}}. 
\eeq 
Indeed, for an even-even axial nucleus, $D_K$ and $D_{-K}$ yields an identical
transition strength.


\section{Results and Discussions} \label{sec:res}
\subsection{Benchmark calculation}
The HFB calculations were done by using SkM* Skyrme parameterization at the ph-channel \cite{SKMS}. 
This set of parameters has been confirmed to be stable in the linear response calculation for the 
infinite nuclear matter~\cite{2012Pastore}. Since SkM* lacks tensor terms, corresponding time-odd terms were also excluded.
For the pp-channel, 
the pairing strengths for neutrons and protons were 
adjusted to reproduce empirical pairing gaps of $^{156}$Dy: 
$V^{\rm pair}_{\rm n}=-282.0$ MeV fm$^3$, $V^{\rm pair}_{\rm p}=-307.9$ MeV fm$^3$. 
The pairing cutoff window needed for the DDDP is fixed to $60$~MeV. 
We use computer code \textsc{hfbtho}, which is an HFB solver based on the harmonic oscillator (HO) basis 
within the axial symmetry \cite{HFBTHO2}. In this work, the HO basis consisted of $20$ shells. 
The imaginary part of $\omega_{\gamma}$ for the FAM strength was set to $\gamma=0.5$~MeV, corresponding
to smearing width of $\Gamma=1.0$~MeV, unless otherwise stated.

We would like to emphasize that, in contrast to the standard solution of the QRPA by matrix diagonalization (MQRPA), 
no truncations on the QRPA quasiparticle model space are imposed in our FAM-QRPA scheme. 
The only cutoffs employed are the number of HO shells and the pairing window, thus, self-consistency between the HFB and QRPA is fully maintained.
\begin{figure}[tb] \begin{center}
  \includegraphics[width=\hsize]{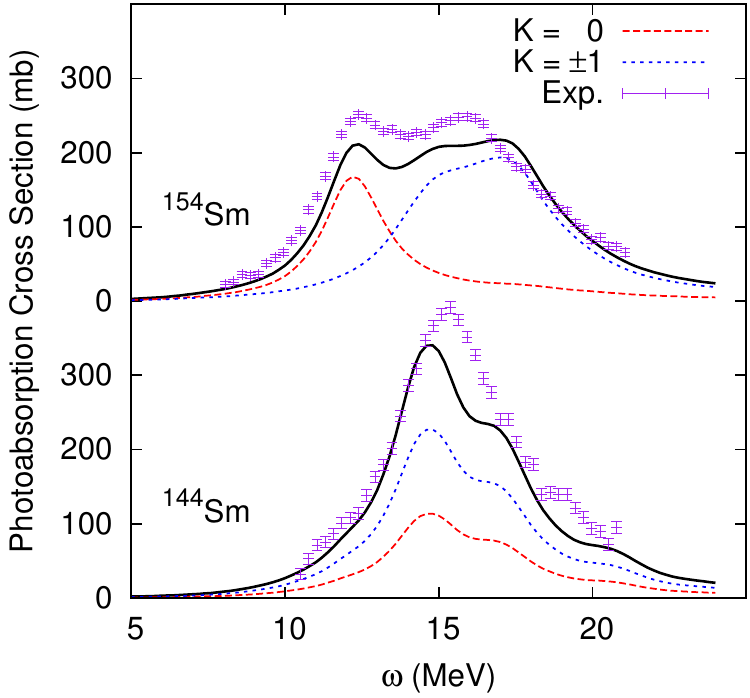}
  \caption{(Color online) Photoabsorption cross sections of $^{144,154}$Sm from the FAM-QRPA calculation. 
The components from the $K=0$ and $K=\pm 1$ modes are also plotted. Here, $K=\pm 1$ refers to
a sum of $+1$ and $-1$ modes. The experimental data is taken from Ref.~\cite{1974Carl}. }
\label{fig:Sm}
\end{center} \end{figure}

Figure \ref{fig:Sm} shows a benchmark result of FAM-QRPA applied to the GDR. 
Here we plotted the photoabsorption cross sections obtained with the IVD strengths 
for $^{144,154}$Sm: 
\beq
 \sigma_{\rm abs}(\omega) = \frac{4\pi^2}{\hbar c} \omega \sum_{K=0,\pm1} \frac{dB(\oprt{D}_K;\omega)}{d\omega}. 
\eeq
The quadrupole matter distribution deformations of the HFB ground states 
are $\beta=0$ and $0.317$ for $^{144}$Sm and $^{154}$Sm, respectively. 
The IVD strengths of the $K=0$ and $\abs{K}=1$ modes split in $^{154}$Sm
due to the ground state deformation, whereas those are identical for spherical $^{144}$Sm.
In both cases, FAM-QRPA shows a good agreement with the experimental data \cite{1974Carl}: 
the typical frequency and width of the GDR can be well reproduced with our model parameters, with the smearing 
width of $\Gamma=1.0$~MeV. 
The plateau distribution for the deformed $^{154}$Sm can be understood as a 
product of the split between $K=0$ and $\abs{K}=1$ modes \cite{2011Yoshida,2013Yoshida}. 
Our result also shows a good consistency to that of Ref.~\cite{2011Yoshida}, 
in which MQRPA within the coordinate-space representation was adopted. 

We have computed FAM strength function within the MPI parallelized scheme,
where each part of the strength function was distributed on a separate core,
similarly as in Ref.~\cite{PhysRevC.92.051302}. 
This scheme achieves a remarkable efficiency, enabling us to compute deformed heavier systems, 
where MQRPA is available only with a truncation of the model space. 
Typically, a computation of both of the $K$-modes took about 1500 CPU hours in a
multicore Intel Sandy Bridge 2.6-GHz processor system.

\begin{table}[tb] \begin{center}
\caption{Ground state properties of Gd, Dy, and Er isotopes obtained with SkM* parameterization: 
axial deformation $\beta$, 
pairing gaps for neutrons and protons $(\Delta_{\rm n},\Delta_{\rm p})$, 
energy-weighted sum rule from Eq.~(\ref{eq:EWSR3}), $m_1(\oprt{D}_K)$, and 
its enhancement factor from the TRK sum rule $\kappa^{\rm IVD}$
(For SkM*, $\kappa^{\rm NM}=0.5315$). }
  \catcode`? = \active \def?{\phantom{0}} 
  \begingroup \renewcommand{\arraystretch}{1.2}
  \begin{tabular*}{\hsize} { @{\extracolsep{\fill}} ccccc} \hline \hline
    Nuclide & $\beta$ & $\Delta_{\rm n},~\Delta_{\rm p}$ & $m_1(\oprt{D}_K)$ & $\kappa^{\rm IVD}$ \\
            &         & [MeV]                          &  [$e^2$fm$^2$MeV] &                   \\ \hline
 $^{152}$Gd    & $0.235$ & $1.09,~1.19$ & $253.9$ & $0.3939$ \\
    $^{154}$Gd & $0.301$ & $1.16,~0.97$ & $256.5$ & $0.3944$ \\
    $^{156}$Gd & $0.331$ & $1.08,~0.89$ & $258.9$ & $0.3950$ \\
    $^{158}$Gd & $0.346$ & $1.05,~0.84$ & $261.3$ & $0.3953$ \\
    $^{160}$Gd & $0.355$ & $1.04,~0.80$ & $263.6$ & $0.3955$ \\
    $^{162}$Gd & $0.358$ & $1.07,~0.78$ & $265.8$ & $0.3955$ \\
    $^{164}$Gd & $0.358$ & $1.06,~0.76$ & $267.9$ & $0.3954$ \\ \hline

 $^{156}$Dy    & $0.289$ & $1.17,~0.98$ & $261.1$ & $0.3944$ \\
    $^{158}$Dy & $0.320$ & $1.13,~0.88$ & $263.7$ & $0.3949$ \\
    $^{160}$Dy & $0.336$ & $1.10,~0.82$ & $266.1$ & $0.3954$ \\
    $^{162}$Dy & $0.344$ & $1.09,~0.78$ & $268.5$ & $0.3956$ \\
    $^{164}$Dy & $0.347$ & $1.09,~0.73$ & $270.8$ & $0.3957$ \\
    $^{166}$Dy & $0.349$ & $1.06,~0.69$ & $273.0$ & $0.3958$ \\
    $^{168}$Dy & $0.348$ & $1.01,~0.67$ & $275.2$ & $0.3959$ \\ \hline

 $^{162}$Er    & $0.324$ & $1.15,~0.87$ & $270.8$ & $0.3953$ \\
    $^{164}$Er & $0.334$ & $1.14,~0.81$ & $273.3$ & $0.3957$ \\
    $^{166}$Er & $0.339$ & $1.12,~0.75$ & $275.7$ & $0.3959$ \\
    $^{168}$Er & $0.342$ & $1.08,~0.70$ & $278.1$ & $0.3962$ \\
    $^{170}$Er & $0.342$ & $1.00,~0.65$ & $280.4$ & $0.3964$ \\
    $^{172}$Er & $0.337$ & $0.97,~0.63$ & $282.6$ & $0.3965$ \\
    $^{174}$Er & $0.329$ & $1.06,~0.61$ & $284.7$ & $0.3964$ \\ \hline \hline
  \end{tabular*}
  \endgroup
  \catcode`? = 12 
\label{tab:gs1}
\end{center} \end{table}

\begin{table}[tb] \begin{center}
\caption{The same as Table \ref{tab:gs1} but for Yb, Hf, and W isotopes. }
  \catcode`? = \active \def?{\phantom{0}} 
  \begingroup \renewcommand{\arraystretch}{1.2}
  \begin{tabular*}{\hsize} { @{\extracolsep{\fill}} ccccc} \hline \hline
    Nuclide & $\beta$ & $\Delta_{\rm n},~\Delta_{\rm p}$ & $m_1(\oprt{D}_K)$ & $\kappa^{\rm IVD}$ \\
            &         & [MeV]                          &  [$e^2$fm$^2$MeV] &                   \\ \hline
 $^{168}$Yb    & $0.331$ & $1.16,~0.60$ & $280.5$ & $0.3961$ \\
    $^{170}$Yb & $0.335$ & $1.10,~0.37$ & $283.0$ & $0.3966$ \\
    $^{172}$Yb & $0.336$ & $1.01,~0???$ & $285.4$ & $0.3970$ \\
    $^{174}$Yb & $0.332$ & $0.94,~0???$ & $287.7$ & $0.3973$ \\
    $^{176}$Yb & $0.324$ & $1.01,~0???$ & $289.9$ & $0.3972$ \\
    $^{178}$Yb & $0.315$ & $1.08,~0???$ & $292.0$ & $0.3970$ \\ \hline

 $^{174}$Hf    & $0.326$ & $1.00,~0.85$ & $290.1$ & $0.3965$ \\
    $^{176}$Hf & $0.316$ & $0.97,~0.80$ & $292.5$ & $0.3970$ \\
    $^{178}$Hf & $0.301$ & $1.03,~0.74$ & $294.8$ & $0.3971$ \\
    $^{180}$Hf & $0.288$ & $1.06,~0.68$ & $297.1$ & $0.3970$ \\
    $^{182}$Hf & $0.276$ & $1.06,~0.64$ & $299.2$ & $0.3969$ \\
    $^{184}$Hf & $0.263$ & $1.03,~0.62$ & $301.4$ & $0.3969$ \\ \hline

 $^{180}$W?    & $0.270$ & $1.09,~0.80$ & $299.7$ & $0.3971$ \\
    $^{182}$W? & $0.257$ & $1.11,~0.73$ & $302.0$ & $0.3972$ \\
    $^{184}$W? & $0.245$ & $1.10,~0.68$ & $304.3$ & $0.3972$ \\
    $^{186}$W? & $0.230$ & $1.07,~0.66$ & $306.5$ & $0.3973$ \\
    $^{188}$W? & $0.212$ & $1.01,~0.65$ & $308.7$ & $0.3974$ \\
    $^{190}$W? & $0.191$ & $0.97,~0.66$ & $310.9$ & $0.3974$ \\ \hline \hline
  \end{tabular*}
  \endgroup
  \catcode`? = 12 
\label{tab:gs2}
\end{center} \end{table}

\subsection{GDR in heavy rare-earth nuclei}
Our survey of the GDR has been performed for even-even rare-earth nuclei 
from Gd ($Z=64$) to W ($Z=74$) isochains. 
Because several sets of experimental data are available \cite{1981Gurevich,1969Berman,1976Gory}, 
they can provide a more systematic check for FAM-QRPA GDR results.
In Tables \ref{tab:gs1} and \ref{tab:gs2}, we summarize the ground state properties of computed nuclei. 
The HFB calculation with SkM* concludes that all the nuclei considered here have 
rather stable prolate deformation. 

\begin{figure}[tb] \begin{center}
  \includegraphics[width=\hsize]{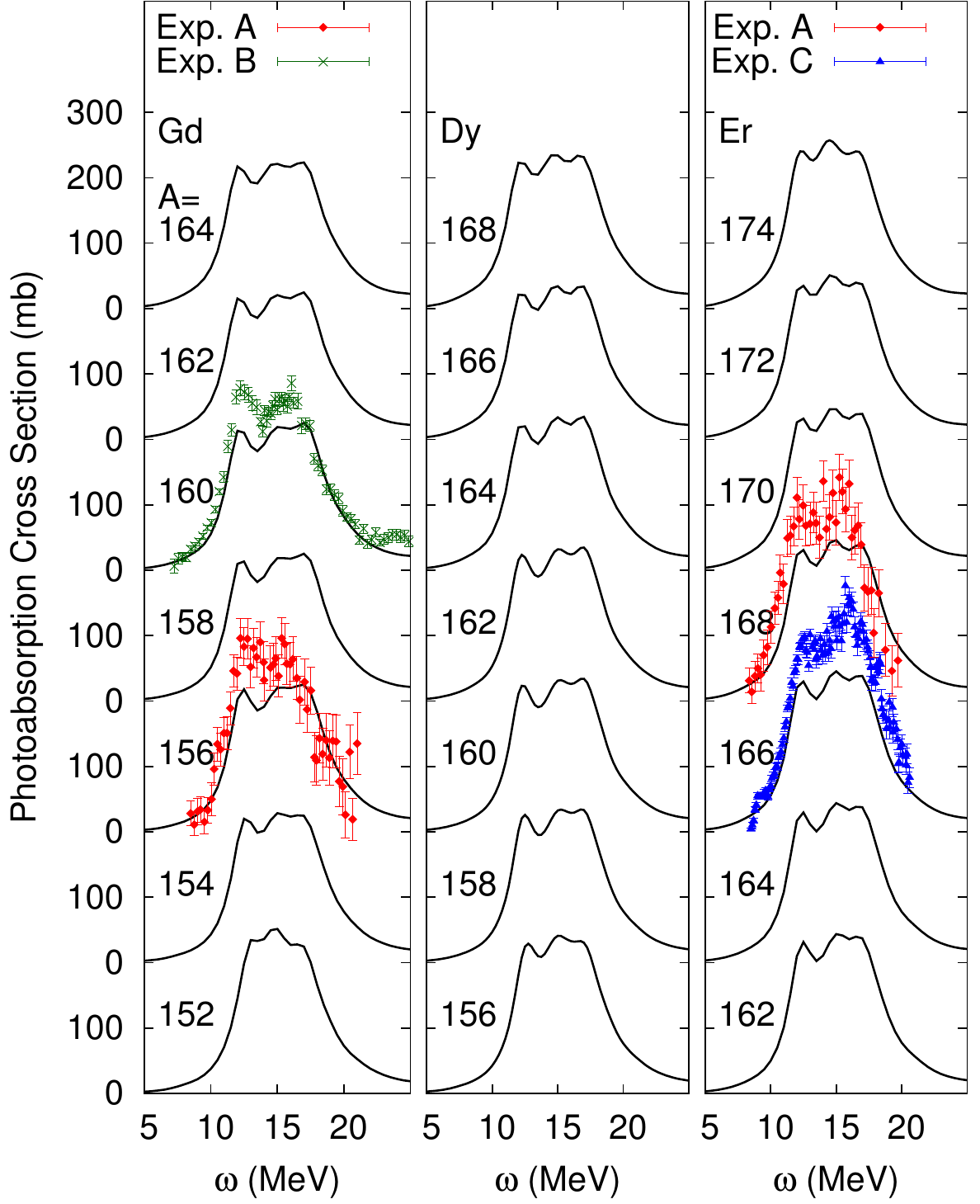}
  \caption{(Color online) 
Photoabsorption cross sections of Gd, Dy, and Er isotopes 
as a function of photon energy. 
For FAM-QRPA calculation, the smearing width $\Gamma=1.0$~MeV is used. 
The experimental data sets A, B (photoabsorption), and C (neutron yield) are taken from 
Refs.~\cite{1981Gurevich}, \cite{1969Berman}, and \cite{1976Gory}, respectively.}
\label{fig:2p0}
\end{center} \end{figure}

\begin{figure}[h] \begin{center}
  \includegraphics[width=\hsize]{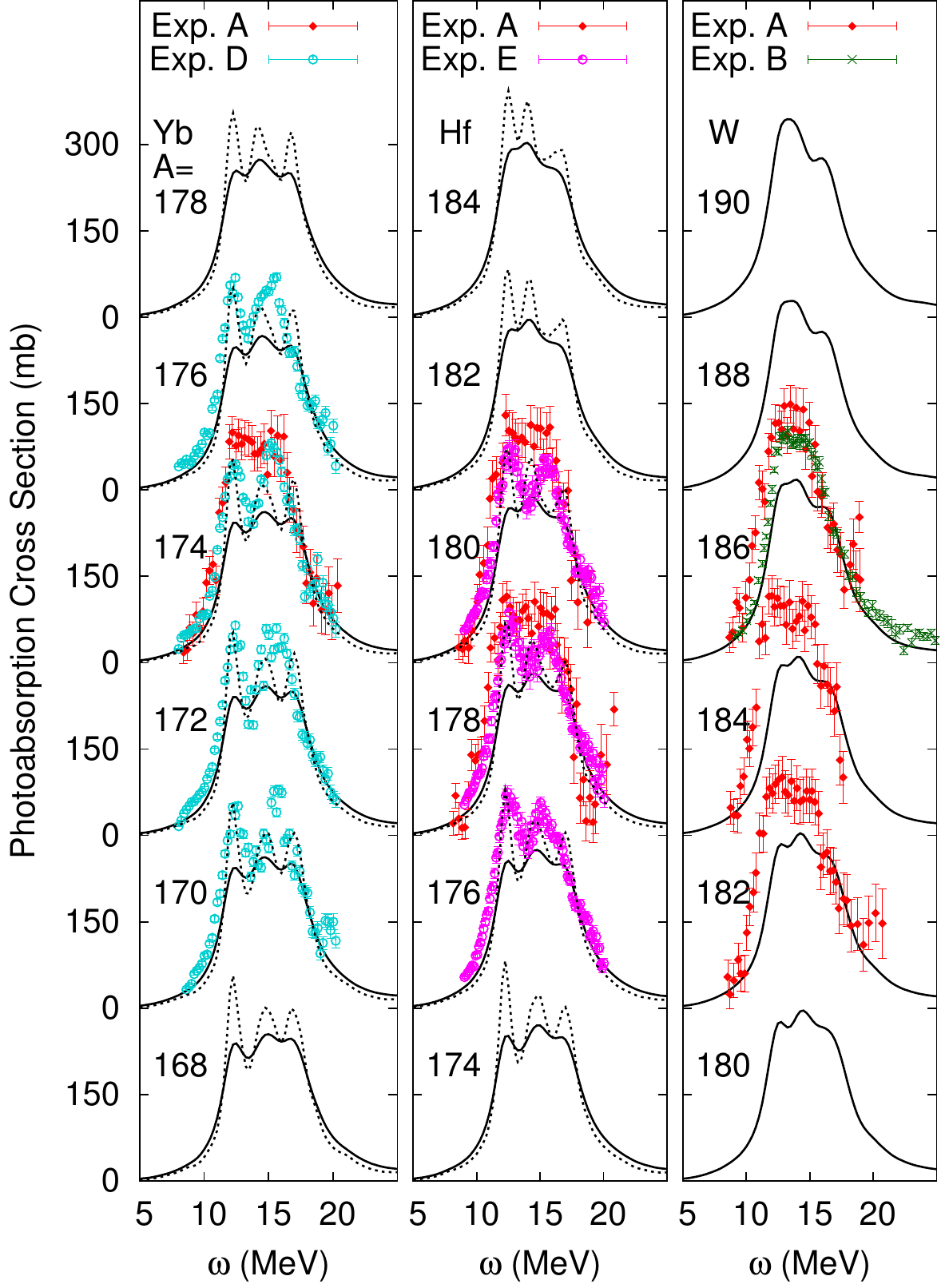}
  \caption{(Color online) 
The same as Fig.~\ref{fig:2p0} but for Yb, Hf, and W isotopes. 
The experimental data sets A, B(photoabsorption),
D(neutron yield), and E(neutron product) are taken from 
Refs.~\cite{1981Gurevich}, \cite{1969Berman},
\cite{1976Gory2}, and \cite{1977Gory}, respectively. 
For Yb and Hf, the results obtained with the smearing width of 
$\Gamma = 0.5$ MeV are also plotted with dotted lines. 
}
\label{fig:2p1}
\end{center} \end{figure}

\begin{figure}[htb] \begin{center}
  \includegraphics[width=\hsize]{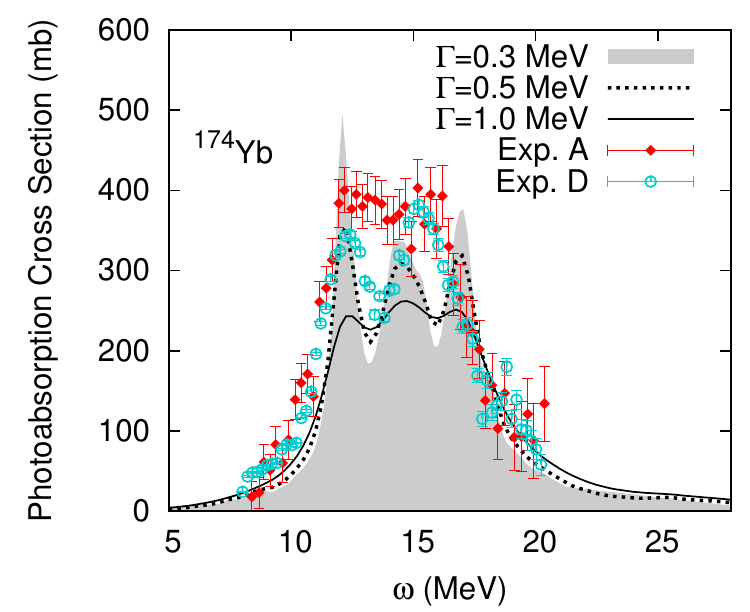}
  \caption{(Color online) 
Photoabsorption cross sections of $^{174}$Yb 
obtained with different values of the smearing width $\Gamma$. 
The shown experimental data is the same as those in Fig.~\ref{fig:2p1}. 
}
\label{fig:174Yb_GX}
\end{center} \end{figure}

\begin{figure}[htb] \begin{center}
  \includegraphics[width=\hsize]{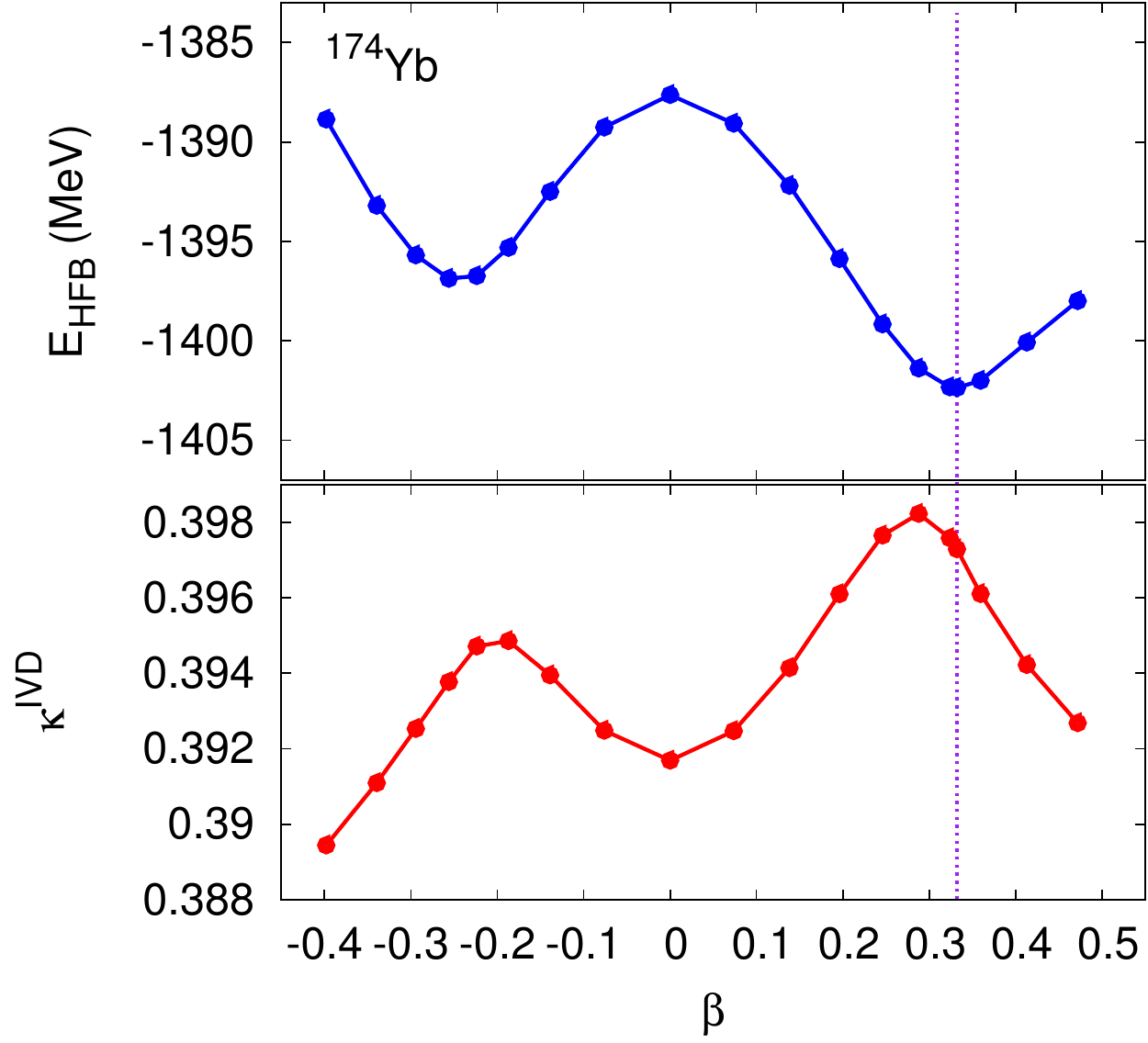}
  \caption{(Color online) The total HFB energy and the IVD enhancement factor of $^{174}$Yb 
as a function of the axial deformation. 
The HFB ground state is indicated by the vertical dotted line. }
\label{fig:beta_kappa}
\end{center} \end{figure}

Our results from FAM-QRPA are summarized in Figs.~\ref{fig:2p0} and \ref{fig:2p1}, in which 
the photoabsorption cross sections are compared with experimental data, where available. 
We emphasize here that experimental data in Refs.~\cite{1976Gory,1976Gory2,1977Gory} does not correspond to
total photoabsorption cross section, but the photo-neutron cross sections of $(\gamma,n)$ and $(\gamma,2n)$ 
type of reactions. Thus, it includes only a part of total photoabsorption cross section due to smaller number of
output channels.
Generally, we find a reasonable agreement between the FAM-QRPA and the experiments. 
Typical frequencies of GDR are fairly well reproduced throughout the rare-earth isotopes heavier than Sm. 
The width and the plateau top of the distribution are well 
understood as a product from the splitting of $K=0$ and $\abs{K}=1$ modes, 
corresponding to the prolate deformation commonly found on their ground states. 
For $^{152}$Gd and several isotopes of W, the width of the GDR is graphically narrower than 
other nuclei, as expected due to smaller prolate deformation. 
In our HFB calculations, the proton pairing collapses for $^{172-178}$Yb. 
This collapse itself, however, does not make a significant impact on the GDR,
as the GDR strength distributions look similar irrespective of the proton pairing collapse.
Although the pairing could affect the GDR indirectly through the ground state properties (mainly deformation), 
their changes are small among the rare-earth nuclei, as shown in the present study.

There is an observed deficiency in the calculated photoabsorption cross sections at 
the region of heavier rare-earth isotopes, namely for $Z\ge 70$. 
For example, the calculated photoabsorption cross section of $^{174}$Yb 
underestimates the experimental data of Ref.~\cite{1981Gurevich} at region $\omega=12-17$~MeV
in which GDR becomes noticeably strong. 
A similar kind of GDR deficiency with Skyrme EDFs was reported in Ref.~\cite{2011Stetcu}. 

In Fig.~\ref{fig:2p1}, we have also plotted results for the Yb and Hf isotopes by
using a smaller smearing width: $\Gamma=0.5$~MeV. 
With this smaller width, we can reproduce cross section of the neutron yield
up to $\omega \approx 13$~MeV, which includes the first peak of the experimental data of Refs.~\cite{1976Gory2,1977Gory}. 
The second peak of the neutron yield cross section may be  attributable to the opening of two-neutron emission channel: 
for $^{174}$Yb, for example, that peak locates at $\omega=15$~MeV, which is just above the two-neutron separation energy. 
With $\Gamma=0.5$~MeV, however, we could not achieve a complete improvement of aforementioned deficiency, 
and the total photoabsorption cross section remains underestimated. In Fig.~\ref{fig:174Yb_GX}, we also show that 
the narrower width of $\Gamma=0.3$~MeV leads computed photoabsorption cross sections 
to overshoot the experimental values at the peak positions, whereas a discrepancy at other frequencies remains. 
Consequently, the GDR deficiency found here is not improved by simply changing the 
smearing parameter $\Gamma$. Further systematic experiments of photoabsorption cross section, 
with improved accuracy, would be helpful by providing a more complete testing ground
for theoretical models.

Theoretical deficiency of the GDR may be connected to the essential properties of the model. 
In order to remedy this deficiency, one could consider {\it e.g.}
beyond QRPA effects or systematic adjustment of the EDF parameters. 
These are, however, 
beyond the scope of this article. 
Alternatively, we discuss a role of the TRK sum rule enhancement factor $\kappa^{\rm IVD}$, 
and its rule in the isovector dipole excitation \cite{1995Rein,1999Rein}. This quantity 
can be related to the isovector effective mass of the infinite nuclear matter (INM), which can 
be used as an input parameter to define the Skyrme EDF parameters \cite{2010Markus}. 
Because there has been some ambiguity about the empirical value of this parameter, 
knowledge of its effect on GDR will be also profitable for the future optimization of the EDF 
parameters.

\subsection{Energy-weighted sum rule}
To discuss the sensitivity of GDR to the model parameters, we investigate the energy-weighted 
sum rule (EWSR), defined as
\beq
m_1(\oprt{D}_K) = \int \omega \frac{dB(\oprt{D}_K;\omega)}{d\omega} d\omega \,.
\label{eq:EWSR1}
\eeq
In terms of the transition matrix elements, it can be rewritten as 
\beq
 m_1(\oprt{D}_K) = \sum_{i>0} \Omega_i \abs{\Braket{i \mid \oprt{D}_K \mid 0}}^2 \,. \label{eq:EWSR2}
\eeq
It is well known that, by applying the Thouless theorem \cite{1961Thouless}, 
the EWSR based on the QRPA can be replaced with the expectation value of 
the double commutator of the HFB ground state \cite{1973MdP,2002Khan}. 
For the present case this reads as
\beqa
 m_1(\oprt{D}_K) &=& \frac{1}{2} \Braket{ 0 \mid [\oprt{D}_K,~[\hat{H},\oprt{D}_K]~] \mid 0} \nonumber \\
 &=& (1+\kappa^{\rm IVD}) \frac{e^2\hbar^2}{2m} \frac{NZ}{A} \frac{3}{4 \pi}, \label{eq:EWSR3} 
\eeqa
where $\kappa^{\rm IVD}$ is the enhancement factor due to the momentum dependence of the effective interaction. 
For the Skyrme force, 
it can be given as 
\beq
 \label{eq:EF_IVD} \kappa^{\rm IVD} = \frac{2m}{\hbar^2} (C^{\rho \tau}_0 - C^{\rho \tau}_1) \frac{A}{NZ}
 \int \rho_{\rm n}(\bir)\rho_{\rm p}(\bir) d\bir . 
\eeq
For the IVD mode, 
the EWSR has the same value for $K=0$ and $1$ cases, even if the ground state is deformed. 
Note also that, for INM, $\kappa^{\rm NM}=2m(C^{\rho \tau}_0 - C^{\rho \tau}_1)\rho_c/\hbar^2$ 
is obtained. 

Before going to applications, we check the validity of Eq.~(\ref{eq:EWSR3}) in a generalized EDF 
framework \cite{2015Hino}. When the EDF is formally generalized, and has no correspondence 
with respect of the underlying effective force, Thouless theorem is not guaranteed to remain valid. 
Because we employed the Skyrme EDF combined with the mixed DDDP, the EWSR from actual QRPA calculations 
can deviate from Eq.~(\ref{eq:EWSR3}). 
In Ref.~\cite{2015Hino}, the authors showed that the Thouless theorem provides a reasonable 
approximation to the EWSR of the isoscalar/isovector monopole and quadrupole modes, 
even when Skyrme EDF lacks exact correspondence with respect to the underlying effective interaction 
but still holds the local gauge invariance. Here we give a similar test for the IVD mode. 

For $^{174}$Yb, the energy-weighted sum rule, integrated from the transition strength function up 
to $\omega=50$~MeV, yields $m_1(\oprt{D}_K)=283.9$ and $283.8$~$e^2$\,fm$^2$\,MeV for $K=0$ and 
$K=1$ modes, respectively. Because of the deformation and the resultant splitting of $K=0$ and $K=1$ strengths, 
there is a small difference between two values. 
The contour integration technique of the complex-energy FAM, developed as an efficient tool to compute the sum rules in Ref.~\cite{2015Hino}, yields $m_1(\oprt{D}_{K=0})=290.1$~$e^2$\,fm$^2$\,MeV with an integration contour radius of $200$̃~MeV.
On the other side, the double commutator procedure of Eq.~(\ref{eq:EWSR3}) 
gives $m_1(\oprt{D}_{K=0, \pm 1})=287.7$~$e^2$\,fm$^2$\,MeV,
with enhancement factor of $\kappa^{\rm IVD}=0.397$, when using HFB proton and neutron densities. 

We find the value from the double commutator method being consistent to those from the FAM-QRPA calculation.
Consequently, Thouless theorem can provide a reliable approximation of the IVD sum rule. 
The computational cost for the double commutator procedure is drastically lighter compared to the FAM-QRPA, 
since it requires information only about the HFB state.

We have summarized the EWSR values and enhancement factors of Eqs.~(\ref{eq:EWSR3}) and (\ref{eq:EF_IVD})
in Tables \ref{tab:gs1} and \ref{tab:gs2}. It is clearly shown that the enhancement factor for the 
TRK sum rule is almost constant at this region of the nuclear chart.
The ratio of two enhancement factors, 
\beq
\label{eq:KR} \frac{\kappa^{\rm IVD}}{\kappa^{\rm NM}} =
\frac{A}{NZ \rho_c} \int \rho_{\rm n}(\bir) \rho_{\rm p}(\bir) d\bir \,,
\eeq
is approximately $0.74$ for rare-earth systems calculated here with SkM* Skyrme parameterization. 
This is simply due to similar value obtained from the density integration 
of Eq.~(\ref{eq:KR}), with only a limited variation on the proton and neutron density profiles 
with respect of $N$ and $Z$.

In Fig.~\ref{fig:beta_kappa}, we plot the enhancement factor $\kappa^{\rm IVD}$ 
and the total HFB energy $E_{\rm HFB}$ as a function of the axial deformation parameter $\beta$ for $^{174}$Yb.
These are obtained from the HFB calculation with a constraint on $\beta$. 
Approximately, the minimum of the total HFB energy corresponds to the maximum of $\kappa^{\rm IVD}$. 
This can be understood mainly in terms of the symmetry energy, which favors a large 
overlap between neutrons and protons. Due to other ingredients, especially the Coulomb energy, 
the true ground state and maximum of $\kappa^{\rm IVD}$ do not exactly coincide.
This curve shows, however, that the IVD enhancement factor is not very sensitive to the 
details of the ground state deformation. 
Next, we investigate if the deficiency against the experimental photoabsorption data found in $^{174}$Yb 
could be improved by changing  $\kappa^{\rm NM}$ parameter.

\begin{table}[tb] \begin{center}
\caption{The INM TRK enhancement factor $\kappa^{\rm NM}$ and $m^*_v/m$, and
the ground state properties of $^{174}$Yb obtained with various Skyrme EDFs.
Note that $m_1(\oprt{D}_K)$ obtained with the double-commutator procedure, Eq.~(\ref{eq:EWSR3}), 
is the same for the different values of the magnetic quantum number $K$. }
  \catcode`? = \active \def?{\phantom{0}} 
  \begingroup \renewcommand{\arraystretch}{1.2}
\begin{tabular*}{\hsize} { @{\extracolsep{\fill}} ccccc c} \hline \hline
 Param.    & $\kappa^{\rm NM}$ & $m^*_v/m$ & \multicolumn{3}{c}{$^{174}$Yb (g.s.)} \\ \cline{4-6}
          &        &                     & $m_1(\oprt{D}_{K=0,\pm 1})$ & $\kappa^{\rm IVD}$ & $\beta$ \\
          &        &                     & [$e^2$fm$^2$MeV]  &                   &         \\ \hline
 SkM*     & $0.5315$ & $0.653$             & $287.7$         & $0.3973$        & $0.312$ \\
 SV-kap20 & $0.2???$ & $0.834$             & $236.1$         & $0.1466$        & $0.337$ \\
 SV-bas   & $0.4???$ & $0.715$             & $266.5$     & $0.2942$        & $0.336$ \\
 SV-kap60 & $0.6???$ & $0.625$             & $297.1$     & $0.4429$        & $0.319$ \\ \hline \hline
\end{tabular*}
  \endgroup
  \catcode`? = 12
\label{table:31}
\end{center} \end{table}

\subsection{Sensitivity to enhancement factor}
In the remaining part of this section, we investigate the sensitivity of GDR to the EDFs with different 
$\kappa^{\rm NM}$ values. 
The Skyrme parameterizations suited to this purpose can be found in Ref.~\cite{2009Klupfel},
where the authors optimized Skyrme parameters by assuming systematic constraints on various 
INM properties. The unconstrained parameterization, SV-min, was optimized without constraints,
and SV-bas parameterization was a base starting point for INM parameter variation. The
parameterizations with a variation on $\kappa^{\rm NM}$  were introduced as SV-kap60 and SV-kap20. 
Because these constrained parameterizations were otherwise optimized exactly in the same manner as SV-bas, 
we can check the effect of $\kappa^{\rm NM}$ on the GDR in a systematic manner. 
The exact value of $\kappa^{\rm NM}$ as an input parameter for each functional is present in Table \ref{table:31}. 
Note that the larger value of the enhancement factor corresponds to the 
lighter isovector effective mass $(\kappa^{\rm NM}=m/m^\ast_v-1)$ \cite{1999Rein,2009Klupfel}. 
Except for the interaction parameterization employed, the numerical conditions are the same 
as in the previous calculations. 

In Table \ref{table:31}, the EWSR of $^{174}$Yb obtained with SkM*, SV-kap20, SV-bas, and SV-kap60 
are summarized with the corresponding enhancement factors. 
As naturally expected from the definitions of $\kappa^{\rm NM}$ and $\kappa^{\rm IVD}$ \cite{1999Rein,2009Klupfel,2015Hino}, 
the EWSR increases with the enhancement factor of INM, in other words, 
with the reduction of the isovector effective mass. 
The HFB ground states computed with three SV-functionals are similarly deformed. Thus, the density 
integration of Eq.~(\ref{eq:KR}) is also similar for all three functionals,
yielding the common ratio of $\kappa^{\rm IVD}/\kappa^{\rm NM}\cong 0.74$.

\begin{figure}[tb] \begin{center}
  \includegraphics[width=\hsize]{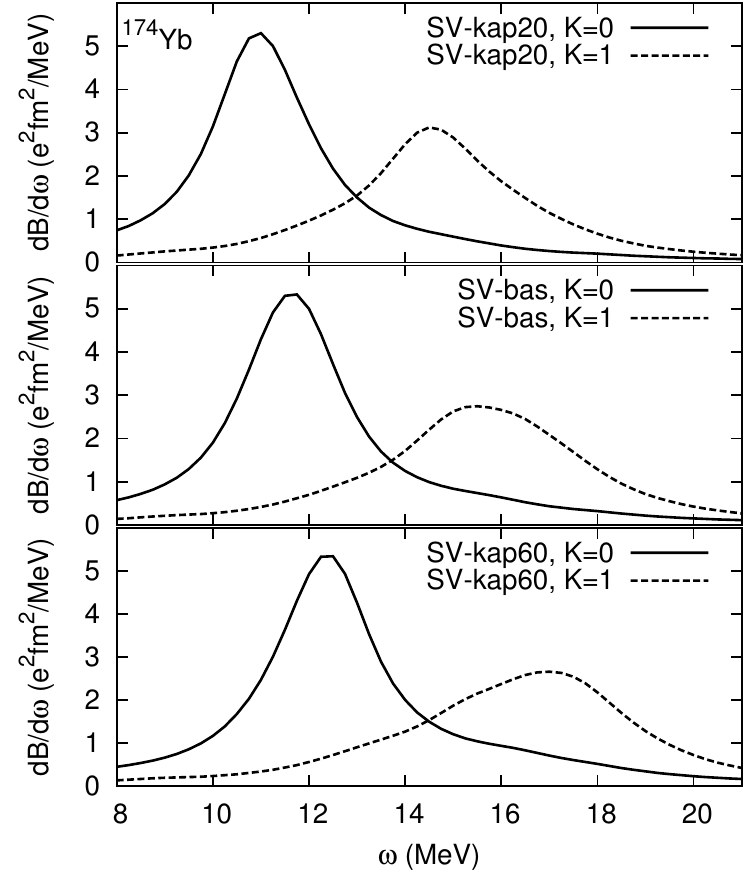}
  \caption{Isovector dipole transition strength in $^{174}$Yb, calculated with 
SV-kap20, SV-bas, and SV-kap60 parameterizations of Ref.~\cite{2009Klupfel}. }
\label{fig:90}
\end{center} \end{figure}

\begin{figure}[h] \begin{center}
  \includegraphics[width=\hsize]{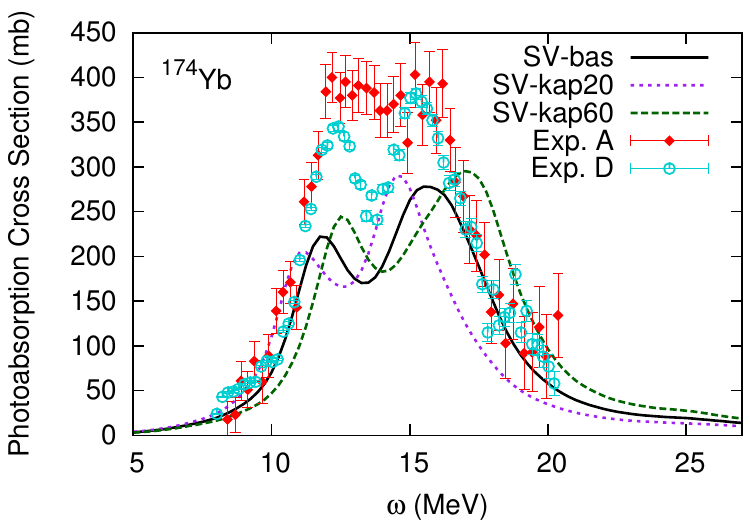}
  \caption{(Color online) 
Photoabsorption cross sections of $^{174}$Yb calculated with 
SV-kap20, SV-bas, and SV-kap60 parameterizations.
The shown experimental data is the same as in Fig.~\ref{fig:2p1}. }
\label{fig:91}
\end{center} \end{figure}

In Fig.~\ref{fig:90}, IVD transition strengths obtained with SV-kap20, SV-bas, and SV-kap60 are plotted. 
One can clearly find that an increment of the enhancement factor leads to a shift of 
the IVD strength towards higher energy region, as pointed out in Ref.~\cite{2009Klupfel} 
for the GDR of doubly-magic $^{208}$Pb.
Conversely, the shifted strength trivially yields an enhanced value due to its energy-weight in the EWSR. 
That is, if the isovector effective mass becomes lighter, the corresponding excitation energies becomes higher. 
This can be qualitatively explained within an analogy to the single-particle energies 
of the HO potential, which have the same curvature, but different particle masses. 
From our calculations, the form of the ground state density is found to be insensitive to $\kappa^{\rm NM}$. 
Thus, in this kind of case, the excitation energies are mainly determined by the 
effective mass for the collective motion.

The energy interval between the $K=0$ and $K=1$ peaks is not sensitive to $\kappa^{\rm NM}$, 
due to similar deformation parameters listed in Table \ref{table:31}. 
It is also noticeable that the total amplitude of the transition strength function is not significantly 
changed for different parameterizations, in both $K=0$ and $K=1$ cases. 
Thus, only the position of the peak is sensitive on $\kappa^{\rm NM}$.
From these results, we expect that the mean frequency of GDR to be a suitable observable 
in order to constrain isovector effective mass parameter during EDF optimization.

The photoabsorption cross sections for SV-EDFs are plotted in Fig.~\ref{fig:91}. 
The shift of GDR structure to the high energy region is again observed
by the increase of $\kappa^{\rm NM}$. 
The SV-kap20 functional reproduces the experimental data well up to $\omega=11$~MeV, 
whereas the general structures matches best to SV-bas result, which corresponds to
$\kappa^{\rm NM}= 0.4$.

On the other side, however, the GDR deficiency against the experimental data
still remains: calculated total photo-absorption cross section cannot be
improved by changing the enhancement factor. 
In order to improve current EDF models, a parameter optimization, combined with an input data
on the GDR position and magnitude, may help the situation.

\section{Summary} \label{sec:sum}
We have performed a systematic analysis of the GDR in heavy rare-earth elements, including 
neutron-rich and deformed isotopes.
The calculations were handled in the recently developed parallel FAM-QRPA scheme, in
combination with the Skyrme EDF,  without any additional truncations on the two-quasiparticle model space. 
This scheme enabled to perform fully self-consistent QRPA calculations efficiently 
and free from the spurious effects due to the broken self-consistency.

The mean energy and width, as well as the plateau shape of the photoabsorption 
cross section, have been fairly well reproduced for nuclei considered. However,
some deficiency in the calculated total photoabsorption cross section was seen
for $Z\geq 70$ isotopes.

We also investigated the behavior of GDR by changing the TRK enhancement factor, 
connected to the isovector effective mass. 
It is clearly shown that an increment of the enhancement factor shifts the GDR 
distribution towards the higher-energy region, corresponding to the lighter isovector effective mass. 

The deficiency of GDR total photoabsorption cross section, with respect to the experimental data,  
remained to be an open question. This deficiency is noticeable at the region of isotopes heavier than Er. 
This situation could not be improved by tuning smearing width or $\kappa^{\rm NM}$. 
Several possibilities of further improvements can be proposed. 
The first is to expand the framework to cover the dynamics beyond the QRPA, as well as the other 
multipole degrees of freedom. 
Especially, the octupole softness of systems could play a role at higher than the
RPA level. 
Another direction of progress is to perform a more systematic optimization of EDF parameters, 
and to use GDR data on deformed nuclei as an input. 
Especially, the tangled effect of the symmetry energy and its slope with 
the isovector effective mass is expected to be important \cite{1999Rein,2008Rein,2012Rein}.

Our FAM-QRPA scheme could be also employed to investigate the low-lying excitations or the pygmy strength. 
For the excitation energies as well as the partial sum rules of these resonances, not only the particle-hole 
part but also the pairing part of the EDF is expected contribute notably. 
Especially, for the low-lying excitations of loosely bound nuclei, 
the HFB method has an advantage over the BCS method for
the treatment of the pairing correlations, especially for the 
nuclei close to the neutron-drip line \cite{1984Jacek,2013Pastore}. 
For this purpose, FAM-QRPA embedded into the coordinate-space HFB solver could be a good choice of the method~\cite{2014Pei}.

\begin{acknowledgments}
We thank Jacek Dobaczewski, Karim Bennaceur, and Andrea Idini for useful comments.
T.O. thanks K. Hagino for interesting discussions. 
This work was supported by Academy of Finland and University of Jyv\"{a}skyl\"{a} 
within the FIDIPRO programme. 
We acknowledge the CSC-IT Center for Science Ltd., Finland and
COMA (PACS-IX) System at the Center for Computational Sciences, University of Tsukuba, Japan, for the allocation of computational resources. 
\end{acknowledgments}


%

\end{document}